\documentclass[english]{elsart}
\usepackage[T1]{fontenc}
\usepackage[latin1]{inputenc}
\usepackage{amsmath}
\usepackage{setspace}
\usepackage{amssymb}
\usepackage{color}
\makeatletter
\providecommand{\LyX}{L\kern-.1667em\lower.25em\hbox{Y}\kern-.125emX\@}
\usepackage{setspace}
\usepackage[numbers]{natbib}
\usepackage[dvips]{graphicx}
\usepackage{epic,epsfig}
\usepackage{subfigure,float}
\usepackage{amsmath,amsfonts,amssymb}
\usepackage{inputenc}
\usepackage{natbib}
\usepackage{euscript}
\makeatother

\begin{document}

\begin{frontmatter}

\title{Spectral and network methods in the analysis of correlation matrices of stock returns}

\author{Tapio Heimo$^{1}$\corauthref{cor1}},
\ead{taheimo@lce.hut.fi}
\author{Jari Saram\"aki$^{1}$},
\author{Jukka-Pekka Onnela$^{1,2}$}, and
\author{Kimmo Kaski$^{1}$}
\corauth[cor1]{Corresponding author.}

\address{$^{1}$ Laboratory of Computational Engineering, Helsinki University of Technology, P.O. Box 9203, FIN-02015 HUT, Finland}

\address{$^{2}$ Physics Department, Clarendon Laboratory, Oxford University, Oxford, OX1 3PU, U.K.}

\begin{abstract}
Correlation matrices inferred from stock return time series contain information on
the behaviour of the market, especially on clusters of highly correlating stocks.
Here we study a subset of New York Stock Exchange (NYSE) traded stocks
and compare three different methods of analysis: i) spectral analysis, \emph{i.e.}
investigation of the eigenvalue-eigenvector pairs of the correlation matrix, ii)
asset trees, obtained by constructing the maximal spanning tree of the correlation
matrix, and iii) asset graphs, which are networks in which the strongest correlations
are depicted as edges.
We illustrate and discuss the localisation of the most significant modes of fluctuation, \emph{i.e.}
eigenvectors corresponding to the largest eigenvalues, on the asset trees and graphs.
\end{abstract}


\end{frontmatter}

\section{Introduction}

The exact nature of interactions between stock market participants is not known but their manifestations in the performance
of stocks are visible. Therefore it is natural to study correlation
matrices of stock returns to learn about the internal structure of the
market. This can be done by
studying the spectral properties of correlation matrices or by
constructing and studying weighted complex networks based on these
matrices (see \emph{e.g.} \cite{mantegna,onnela:dynam,tumminello,tumminello2}
and references therein).
Here, we compare these two approaches.

The paper is organised as follows: in Section 2 we give a short
introduction to financial correlation matrices and their spectral
properties. A comparison of the spectral properties and results obtained using asset trees
and graphs is presented in Section 3. Summary and conclusions are given in Section 4.

\section{Correlation matrix and its spectral properties}

Our dataset consists of the split-adjusted daily closing prices of
$N=116$ stocks, traded at the New York Stock Exchange (NYSE) for the
time period from 13-Jan-1997 to 29-Dec-2000. This amounts to 1000 price
quotes per stock. The equal time correlation matrix of
logarithmic returns is constructed by
\begin{equation} \label{Cormat}
C_{ij} = \frac{\langle G_{i} G_{j} \rangle -  \langle G_{i} \rangle
  \langle G_{j} \rangle}{\sigma_{i} \sigma_{j}},
\end{equation}
where $\sigma_{i} = \sqrt{\langle G_{i}^{2} \rangle - \langle G_{i}
  \rangle^{2}}$, $G_{i}(t) = \ln P_{i}(t) - \ln P_{i}(t-1)$, $P_{i}(t)$ 
is the price of stock $i$ at time $t$ and the angular brackets denote
  time average. From Eq.~\ref{Cormat} we see that
the correlation matrix $C$ is the covariance
  matrix of the time series rescaled to have unit variance. 
These time series can be seen as $T$ realisations of a
random vector $Z$ in $\mathbb{R}^N$, assuming that the elements of the time series are real
numbers and we have $N$ time series of length $T$. By diagonalising $C$ we can find an orthogonal system of coordinates 
where the components of $Z$ do not correlate. These components
are usually called the \emph{principal components}. The elements of
the diagonal matrix, the eigenvalues, implicate the variances of the
corresponding principal components. In the following we denote the
  eigenvectors of $C$ by $x_{1}, \ldots , x_{N}$ and the corresponding
  eigenvalues by $\lambda_{1}, \ldots , \lambda_{N}$, where
  $\lambda_{1} > \ldots > \lambda_{N}$.

The eigenvectors can be thought to represent \emph{modes of fluctuation}.
The time series studied here are such that the rescaling makes them comparable with
each other and this is clearly inherited to the principal
components. Thus the eigenvalues reflect the significance of the
corresponding modes of fluctuation. 

The correlation matrix $C$ of $N$ assets has $N(N-1)/2$ distinct 
entries. Assuming that one determines an empirical correlation matrix
from $N$ time series of length  $T$ and $T$ is not very large compared
to $N$, the entries of the correlation matrix are very noisy and the
matrix is to a large extent random. 
Laloux \emph{et al.}~\cite{laloux}
and Plerou \emph{et al.}~\cite{plerou} have studied the spectral
properties of financial correlation matrices and concluded that only
few eigenpairs carry real information. Their work suggests that the
eigenvalues can be classified as follows:
\begin{enumerate}
\item The very smallest eigenvalues do not belong to the random part
  of the spectrum. The corresponding eigenvectors are highly localized, \emph{i.e.},
  only a few assets contribute to them.

\item The next smallest eigenvalues (about 95 \% of all eigenvalues)
  form the ``bulk'' of the spectrum. They or at least  most of them
  correspond to noise and are well described by random matrix theory.

\item The largest eigenvalue is well separated from the bulk and
  corresponds to the whole market as the corresponding eigenvector has
  roughly equal components.

\item The next largest eigenvectors carry information about the real
  correlations and can be used in identifying clusters of strongly
  interacting assets. 
\end{enumerate}

\section{Asset trees, asset graphs and eigenvector localisation}

\begin{figure}
\begin{center}
\includegraphics[width=0.7\textwidth]{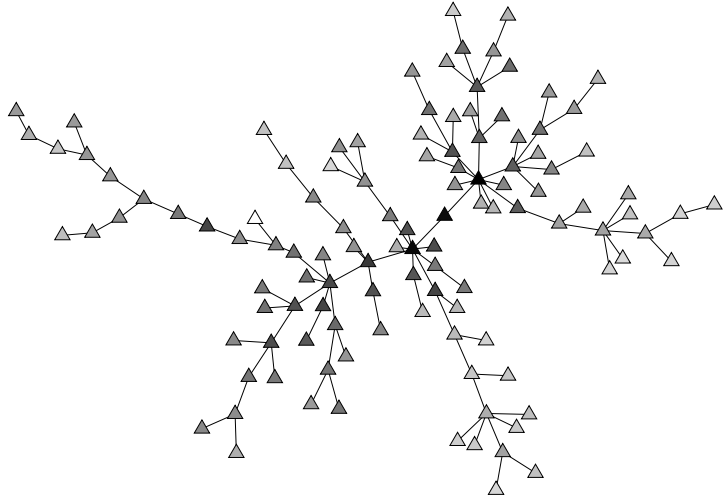}
\caption{The asset tree, displaying the values of the components of the most significant 
mode of fluctuation, the market eigenvector $x_{1}$. 
The color of a node denotes the contribution of the
corresponding component of $x_{1}$ to the length of
the eigenvector. The largest component is colored black. For other
nodes, linear scale is used such that white color indicates zero
contribution. }
\label{fig:mst}
\end{center}      
\end{figure}

In addition to spectral analysis, correlation matrices of stock return time series 
have recently been analyzed with network-related methods. The aim has been to uncover
structure in the correlation matrix in the form of clusters of highly correlating stocks.
In this section we discuss how the eigenvectors corresponding to the largest eigenvalues
are localized with respect to clusters of stocks inferred using the \emph{asset 
tree}~\cite{mantegna,onnela:dynam} and \emph{asset graph}~\cite{onnela:clust} approaches.

The maximal spanning tree of the stocks, later referred to as the asset
tree, is a simply connected graph consisting of all $N$ stocks and $N-1$ edges such that the sum
of the correlation coefficients between the endpoints of each edge is
maximized.  Fig.~\ref{fig:mst} displays the asset tree for our data set, together with the 
market eigenvector $x_{1}$, \emph{i.e.},
the most significant mode of fluctuation.
The color of a node denotes the contribution of the corresponding
eigenvector component to the length of the eigenvector (\emph{i.e.}, the square of the
component). The linear color map is chosen such that white color indicates
zero contribution and the largest component is denoted by black, shades of grey
depicting smaller component values. We see that the most central
nodes of the asset graph contribute most to the
market eigenvector. This is rather natural,
as the central nodes in the asset graph are known to be very
large multisector companies or investment banks, which obviously
fluctuate as their diversified investments~\cite{onnela:dynam}. 
\begin{figure}
\begin{center}
\includegraphics[width=160pt]{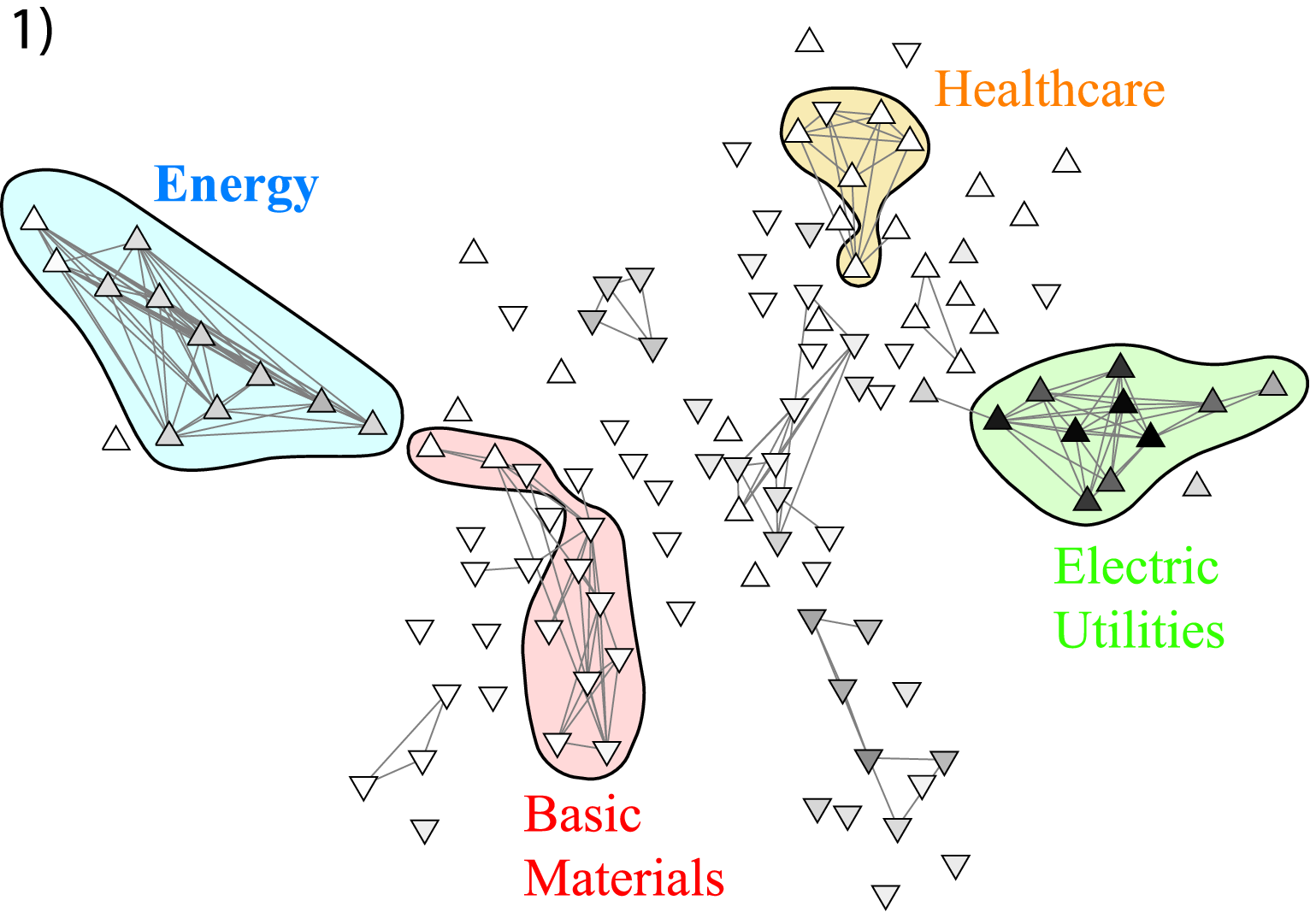}
\includegraphics[width=160pt]{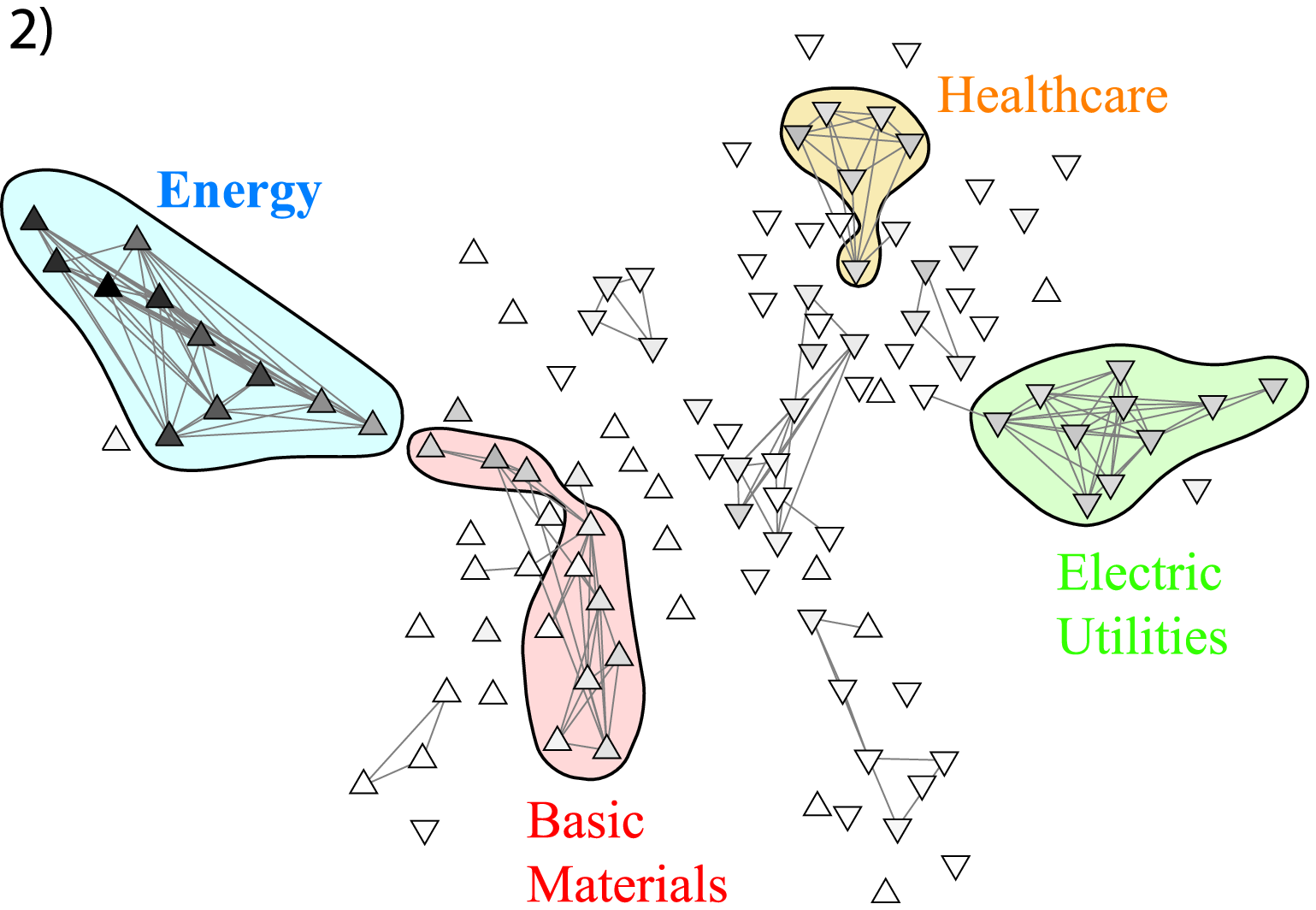}
\includegraphics[width=160pt]{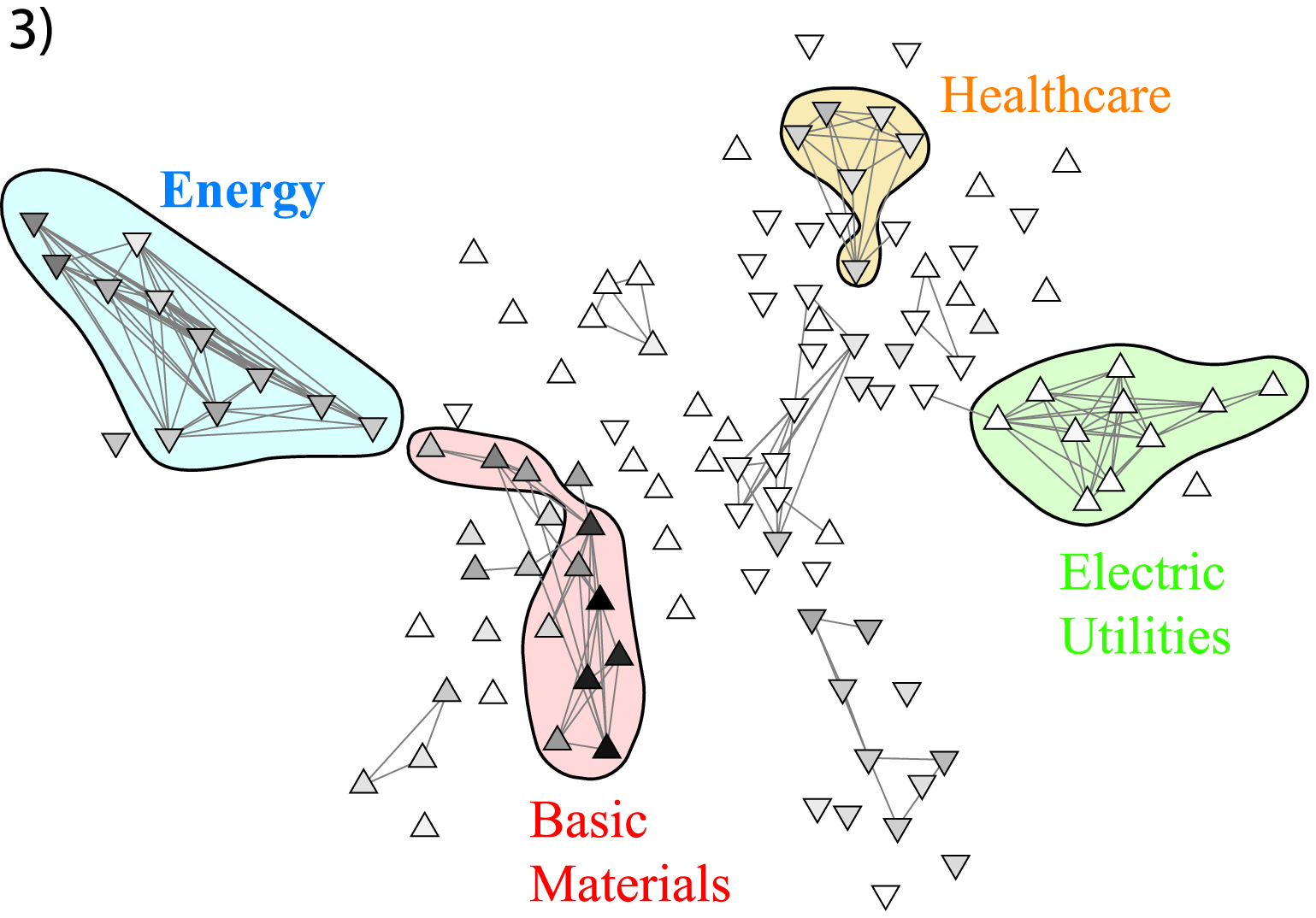}
\includegraphics[width=160pt]{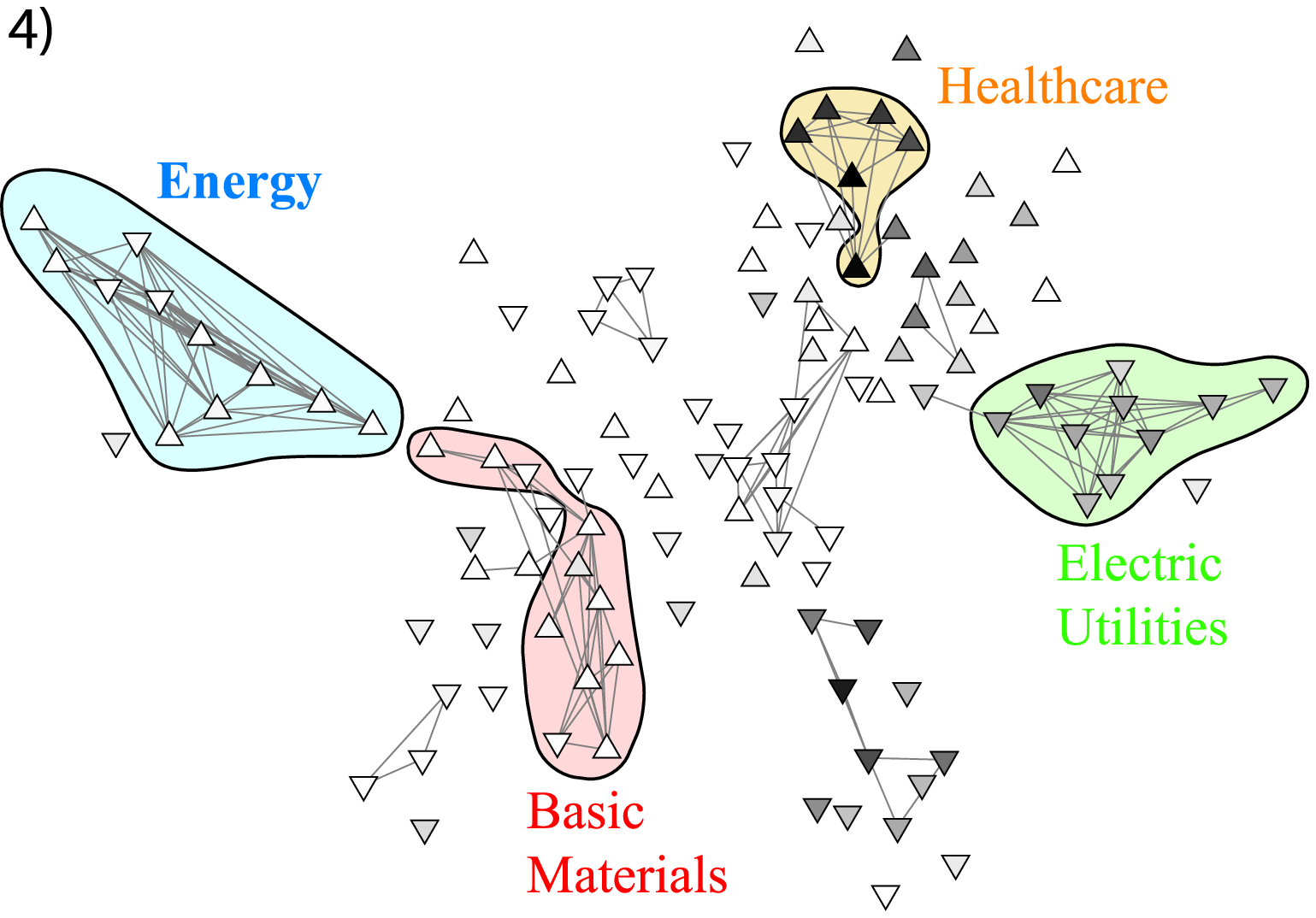}
\caption{The asset graph for occupation $p=0.025$ and the localisation of
$x_{2}$, $x_{3}$, $x_{4}$ and $x_{5}$ (panels 1-4, respectively). The
orientation of the triangle at a node denotes the sign of the corresponding
eigenvector component, and the color is determined as in
Fig.~\ref{fig:mst}. Clusters corresponding to these eigenvectors,
identified by the clique percolation method, are denoted by the shaded background.}
\label{fig:assetpanels}
\end{center}      
\end{figure}

As discussed earlier by Mantegna~\cite{mantegna} and Onnela \emph{et al.}~\cite{onnela:dynam},
the asset tree contains a lot information about the
clustering of stocks. Therefore it is very interesting to compare
the localization of the next most significant modes and the topology of the asset
tree. From Fig.~\ref{fig:assetpanels}, where the nodes are plotted
with the same coordinates as in Fig.~\ref{fig:mst} and the
localisation of $x_{2}$, $x_{3}$, $x_{4}$ and $x_{5}$ is illustrated (panels 1-4, respectively), we see that these
modes are mainly localised to branches of the asset tree. In these
eigenvectors, unlike in the market eigenvector $x_{1}$, components of 
both signs exist. The sign of a component 
is denoted by the orientation (up or down) of the triangle at the corresponding
node. According to the
Forbes classification~\cite{forbes} the above mentioned branches approximately 
correspond to the Electric
Utilities industry of the Utilities sector, the Energy sector, the
Basic Materials sector and the Healthcare sector.

Onnela \emph{et al.}~\cite{onnela:clust} were the first to study the clustering of stocks
using asset graphs constructed from correlation matrices of
returns. An asset graph is constructed by ranking the non-diagonal
elements of the correlation matrix $C$ in decreasing order 
and then adding a set fraction of links between stocks starting from the strongest correlation
coefficient. The emergent network can be characterised by a parameter
$p$, the ratio of the number of added links to the number of all
possible links, $N(N-1)/2$. Evidently, the higher the value of $p$, the denser the
resulting network; in our view the question of whether some specific value
of $p$ yields the most informative structure is still open (see Ref.~\cite{onnela:clust} for
results obtained by sweeping the $p$ value). For the following
analysis we have simply chosen $p=0.025$ as with this value the strongest
clusters are clearly visible. In order to identify the visually apparent cluster
structure, we have utilized the clique percolation method introduced by Palla
\emph{et al.}~\cite{palla}, using cliques of size three. The four clusters detected
with this method, best corresponding to eigenvectors $x_{2}, \ldots, x_{5}$ are illustrated in Fig.~\ref{fig:assetpanels}. 
The clusters are seen to mostly correspond to the above-mentioned
industry sectors. 

\begin{table}
\scriptsize
\begin{center}
\begin{tabular}[]{|l|c|c|c|c|c|}
\hline
 cluster vector $e_{c}$ & $x_{1}$ & $x_{2}$ & $x_{3}$ & $x_{4}$ & $x_{5}$\\
\hline
Electric Utilities & 0.2277 & \textbf{0.7117} & -0.3585         & 0.0708 & -0.4117\\
Energy             & 0.3026 & 0.3148          & \textbf{0.7190} & -0.4299 & 0.0298\\
Basic Materials    & 0.3451 & -0.0739         & 0.2859          & \textbf{0.6096} & 0.0762\\
Healthcare         & 0.2510 & 0.0388          & -0.2691         & -0.2416 & \textbf{0.5042}\\
\hline
\end{tabular}
\caption{The Euclidean inner products of the vectors describing the
  clusters and the eigenvectors $x_{1}, \ldots, x_{5}$. The largest
  value of each row is bolded.} 
\label{dots}
\end{center}
\end{table}

From Fig.~\ref{fig:assetpanels} we see that $x_{2}$ and $x_{3}$ are
rather strongly localised to the respective clusters; however,
$x_{4}$ and especially $x_{5}$ no longer match the clusters
well. The localisation of the market eigenvector $x_{1}$ and the following four
eigenvectors is quantified in Table I, which displays the inner products
of these eigenvectors and vectors depicting the clique percolation clusters. 
We have defined a normalized vector to
depict each cluster such that $e_{c}=[e_{c}^{1}, \ldots, e_{c}^{N}]^{T}$, where $e_{c}^{i}$ is
constant for all components belonging to cluster $c$ and zero for
other components. It is seen that $x_{1}$ is rather 
evenly distributed in the clusters, whereas $x_{2}$ and $x_{3}$ 
are mostly localised on the Electric Utilities and Energy clusters, respectively.
Similarly $x_{4}$ and $x_{5}$ are mostly localised on the Basic Materials
and Healthcare clusters. However, the difference to other clusters appears to become 
smaller with increasing eigenvector index. This is corroborated by analysis
of further eigenvectors (not shown); with some exceptions the 
eigenvectors with higher indices 
appear less well localized with respect to clusters of the asset graph
or branches in the asset tree.

\section{Summary and conclusions}

We have studied and compared how strongly correlated clusters of stocks are
revealed as branches in the asset tree, as clusters in asset graphs, 
and as non-random eigenpairs of the correlation matrix.
The eigenvector corresponding to the largest eigenvalue has roughly
equal components, but the components corresponding to the most
central nodes of the asset tree are on average somewhat larger than
others. The eigenvectors corresponding to the next largest eigenvalues are
to some extent localised on branches of the asset tree. When comparing the
localization of these eigenvectors to clique percolation clusters,
it is seen that the first few eigenvectors match the clusters rather well.
However, their borders are ``fuzzy'' and do not define clear cluster boundaries. 
With increasing eigenvector index, the eigenvectors appear to localize increasingly
less regularly with respect to the asset graph (or asset tree) topology.
Hence it appears that identifying the strongly interacting clusters of stocks solely
based on spectral properties of the correlation matrix is rather difficult; the asset
graph method seems to provide more coherent results.

\textbf{Acknowledgments}

The authors would like to thank János Kertész and Gergely Tibély for
useful discussions. This work has been supported by the Academy of
Finland (the Finnish Center of Excellence program 2006-2011).


\end{document}